%% file: main.tex
\documentclass[sigconf, 10pt]{acmart}

\setcopyright{none}
\renewcommand\footnotetextcopyrightpermission[1]{}

\usepackage{fancyhdr}
\fancypagestyle{plain}{%
   \fancyhf{} %
   \fancyfoot[L]{ACM SIGCOMM Computer Communication Review}%
   \fancyfoot[R]{Volume 48 Issue 5, October 2018}%
}
\fancypagestyle{firstpagestyle}{%
   \fancyhf{} %
   \fancyfoot[L]{ACM SIGCOMM Computer Communication Review}%
   \fancyfoot[R]{Volume 48 Issue 5, October 2018}%
}

\usepackage{booktabs}
\usepackage{syntax}
\usepackage{newfloat}
\usepackage{color}
\usepackage{setspace}
\usepackage{subfigure}
\usepackage{soul}
\usepackage{url}
\usepackage{balance}
\usepackage{hyperref}

\DeclareFloatingEnvironment[
  fileext   = logr,
  listname  = {List of Grammars},
  name      = Grammar,
  placement = htp
]{grammarbox}

\usepackage{listings}
\usepackage{color}
\usepackage{enumitem}

\definecolor{mygreen}{rgb}{0,0.6,0}
\definecolor{mygray}{rgb}{0.5,0.5,0.5}
\definecolor{mymauve}{rgb}{0.58,0,0.82}

\copyrightyear{2018} 
\acmYear{2018} 
\setcopyright{acmcopyright}
\acmConference[SelfDN '18]{ACM SIGCOMM 2018 Workshop on Self-Driving Networks}{August 24, 2018}{Budapest, Hungary}
\acmPrice{15.00}
\acmDOI{10.1145/3229584.3229590}
\acmISBN{978-1-4503-5914-6/18/08}

\everypar{\looseness=-1}

\begin{document}
\title{Refining Network Intents for Self-Driving Networks}

\author{Arthur Selle Jacobs}
\affiliation{%
  \institution{UFRGS}
}
\email{asjacobs@inf.ufrgs.br}

\author{Ricardo Jos\'e Pfitscher}
\affiliation{%
  \institution{UFRGS}
}
\email{rjpfitscher@inf.ufrgs.br}

\author{Ronaldo Alves Ferreira}
\affiliation{%
  \institution{UFMS}
}
\email{raf@facom.ufms.br}

\author{Lisandro Zambenedetti Granville}
\affiliation{%
  \institution{UFRGS}
}
\email{granville@inf.ufrgs.br}

\renewcommand{\shortauthors}{A. S. Jacobs, R. J. Pfitscher, R. A. Ferreira and L. Z. Granville}

\input{src/abstract}

%
%



\maketitle

\let\thefootnote\relax\footnotetext{This is a slightly revised version of the paper 'Refining Network Intents for Self-Driving Networks' that was initially presented at the ACM SIGCOMM'18 Workshop on Self-Driving Networks (SelfDN 2018).}

\input{src/introduction}
\input{src/refinement}

\input{src/language}

\input{src/implementation}

\input{src/evaluation}
\input{src/related}
\input{src/conclusions}

\section*{Acknowledgments}

We thank our shepherd Walter Willinger and the anonymous reviewers for their valuable feedback. This work was supported in part by the Brazilian National Research and Educational Network (RNP), the Brazilian  Federal Agency  for  Support  and  Evaluation  of  Graduate  Education (CAPES), and the Brazilian National Council for Scientific and Technological Development (CNPq). This research is part of the INCT of the Future Internet for Smart Cities funded by CNPq proc. 465446/2014-0, CAPES proc. 88887.136422/2017-00, and FAPESP proc. 14/50937-1 and FAPESP proc. 15/24485-9.

\balance
\bibliographystyle{ACM-Reference-Format}
\bibliography{bibtex}

\end{document}

%% file: src/abstract.tex
\begin{abstract}

Recent advances in artificial intelligence (AI) offer an opportunity for the adoption of self-driving networks. However, network operators or home-network users still do not have the right tools to exploit these new advancements in AI, since they have to rely on low-level languages to specify network policies. Intent-based networking (IBN) allows operators to specify high-level policies that dictate how the network should behave without worrying how they are translated into configuration commands in the network devices. However, the existing research proposals for IBN fail to exploit the knowledge and feedback from the network operator to validate or improve the translation of intents. In this paper, we introduce a novel intent-refinement process that uses machine learning and feedback from the operator to translate the operator's utterances into network configurations. Our refinement process uses a sequence-to-sequence learning model to extract intents from natural language and the feedback from the operator to improve learning. The key insight of our process is an intermediate representation that resembles natural language that is suitable to collect feedback from the operator but is structured enough to facilitate precise translations. Our prototype interacts with a network operator using natural language and translates the operator input to the intermediate representation before translating to SDN rules. Our experimental results show that our process achieves a correlation coefficient squared (\textit{i.e.}, R-squared) of 0.99 for a dataset with 5000 entries and the operator feedback significantly improves the accuracy of our model.

\end{abstract}

%% file: src/introduction.tex
\section{Introduction}
A {\em self-driving network} is an autonomous network that can predict changes and adapt to user behaviors without the intervention of an operator. Successfully implementing an autonomous network would not only ease network management but also reduce operational costs. Recent advances in artificial intelligence (AI) offer an opportunity for the adoption of self-driving networks, as machine learning models can identify patterns and learn how to respond to changes in the network.  However, network operators still do not have the right tools to exploit these new developments in AI, since they still have to rely on low-level languages to specify network policies and complex interfaces to ensure that the specified policies are deployed correctly. Moreover, home-network users do not have the skills to program their networks and can benefit from a friendly management system.

Intent-based networking (IBN) allows operators to specify high-level policies that dictate how the network should behave---\textit{e.g.}, defining goals related to quality of service, security, and performance--without worrying about the low-level details that are necessary to program the network to achieve these goals. Existing research proposals for IBN present several intent languages, frameworks, and compilers to deploy intents in network devices and middleboxes~\cite{Prakash2015, Abhashkumar2017, Sung2016, Ryzhyk2017}. These proposals enable composition of high-level policies~\cite{Prakash2015, Ryzhyk2017}, deployment in software-defined networks (SDN) \cite{Abhashkumar2017}, and management abstractions for network operators~\cite{Sung2016}. While these are steps in the right direction, these proposals cannot extract intent information from pure natural language, requiring that network operators learn a new intent definition language in each proposal and, consequently, hindering interoperability, deployment, and management of heterogeneous networks. 

Most of the existing research proposals for IBN fail to exploit the knowledge and feedback from the network operator.  Highly complex and, sometimes, conflicting policies in network devices may cause network intents to derail from the desired behavior of the operator. Moreover, the adoption of programmable network technologies, such as SDN and Network Functions Virtualization (NFV)~\cite{ETSI-WHITE-PAPER-1}, introduce a new level of dynamism that results in constant changes in network conditions.  Therefore, monitoring the network after deploying policies and requesting feedback from the operator are crucial for avoiding misconfigurations. 

In this paper, we introduce a novel intent-refinement process that uses machine learning and feedback from the operator to translate the operator's utterances into network configurations (\textsection\ref{sec:refinement}). Our process consists of three stages.  First, we rely on an intelligent chatbot interface to extract the main actions and targets (\textit{i.e.,} \textit{entities}) of an user intent from natural language (\textsection\ref{sec:extraction}).  We implement the chatbot interface using DialogFlow~\cite{dialogflow}, which uses machine learning to identify key aspects in the user's utterances without the need for extensively covering every possible entity value.  In our chatbot, examples of entities are the network endpoints, middleboxes, and temporal configurations for the policy. A natural language interface enables the deployment of our solution in distinct scenarios. For instance, a home user could use our chatbot to prioritize streaming traffic in her network during specific hours of the day.

Second, we use a neural sequence-to-sequence learning model to translate the extracted entities into a high-level structured network definition program (\textsection\ref{sec:translation}). The program is written in \textit{Nile}, our new structured intent definition language (\textsection\ref{sec:language}), which closely resembles natural language. The \textit{Nile} program is then presented to the network operator for confirmation on the extracted behavior. For home users with no technical knowledge, the confirmation can come from a voice assistant or a graphical interface.

Finally, we compile the extracted intent program into a network policy according to the destination network (\textsection\ref{sec:deployment}).  As a proof-of-concept, we implement a service chain for specific traffic using SONATA-NFV~\cite{Peuster2016}(\textsection\ref{sec:implementation}). However, the decoupling provided by the intent definition language allows the compilation of the intents to other existing network configurations---including policy languages, such as Janus~\cite{Abhashkumar2017}, PGA~\cite{Prakash2015}, and Kinectic~\cite{Kim2015}---improving the reusability of our proposed solution.  In this stage, we also make assertions to verify any conflicts between the extracted intent and the network configuration---e.g., an intent asking for more bandwidth than is available on the required path---and warn the operator through the chatbot interface.

In summary, our key contributions in this paper are:
\vspace{-1mm}
\begin{enumerate}[leftmargin=*]
\item A novel intent-refinement process for intelligent extraction of intents from natural language that uses feedbacks from network operators to improve learning. 


\item \textit{Nile}: a high-level, comprehensive intent definition language (\textsection\ref{sec:language}) that resembles the English language. Nile acts as an abstraction layer for other policy mechanisms, reducing the need for operators to learn a new policy language for each different type of network.

\item Experimental results that show significant improvements on translation accuracy with the feedback from the operator (\textsection\ref{sec:evaluation}).
\end{enumerate}

%% file: src/refinement.tex
\section{Refinement Process}
\label{sec:refinement}

The first requirement for a self-driving network to reduce its management complexity is intelligent and seamless planning.  A network operator should be able to specify network policies without worrying how they would be achieved. It would be even better if the network operator could use natural language to define the network behavior. The behavior may include customer expectations to comply with Service Level Agreements (SLAs),  network functions for security, temporal behavior for accommodating large flows during peak hours, or network-wide goals like minimizing congestion or reducing traffic costs by relying on cheaper paths in the network.

\begin{figure}[!ht]
	\centering
	\includegraphics[width=\linewidth]{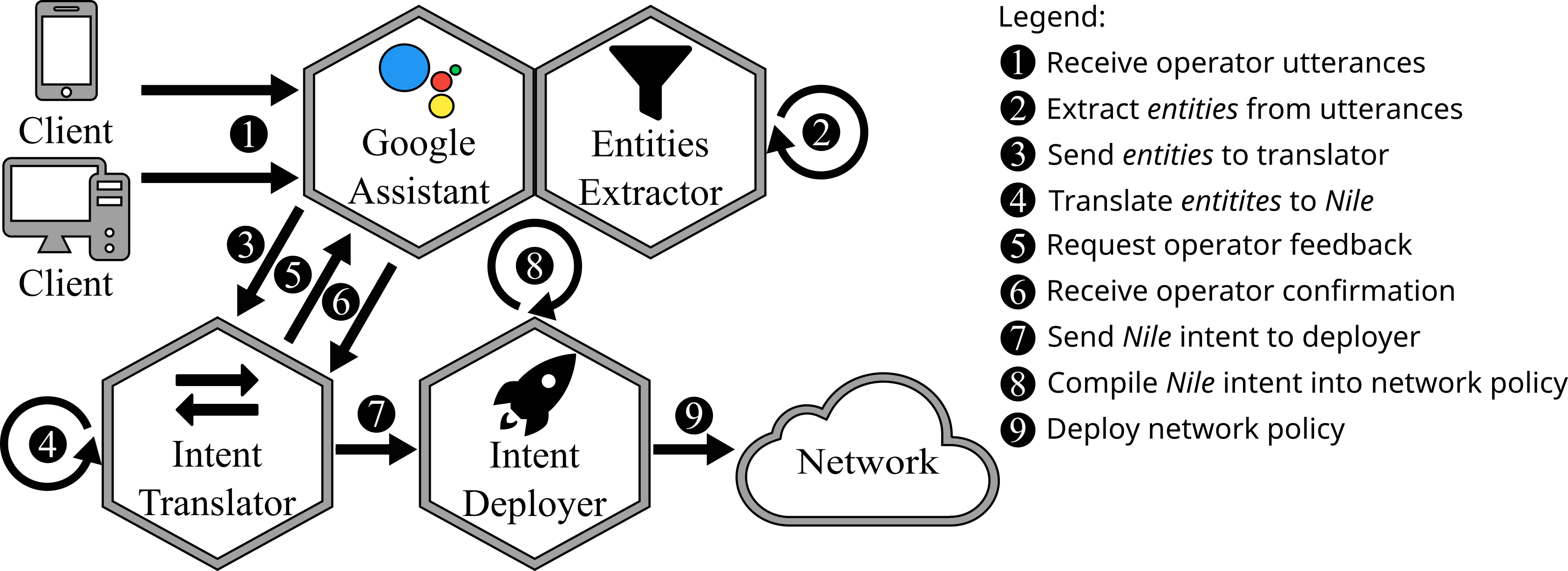}
    \caption{Intent Refinement Process.}
    \label{fig:architecture}
\end{figure}

With the above requirements in mind, we propose a refinement process for intent specification that can learn and adapt itself to achieve the network behavior expressed by the operator while providing a user-friendly interface for interactions with the operator.  This section presents the three stages of the refinement process: entities' extraction, intent translation, and intent deployment. Figure \ref{fig:architecture} presents an overview of the refinement process with the three stages and the steps involved in translating intents described in natural language to network configurations. Note that the operator provides feedback via chatbot interface in Step 6, and Steps 2-6 are repeated until the operator confirms the correct translation of the intents.

\subsection{Entities Extraction}
\label{sec:extraction}

The first step in the intent refinement process is to extract the actions and targets of the network behavior expressed in natural language by the operator. In this step, we use DialogFlow~\cite{dialogflow} to build the Entities Extractor. DialogFlow (formerly known as API.AI) is a development framework to build human-computer interactions based on natural language conversations (\textit{i.e.}, chatbots). The framework uses machine learning to generalize example cases referred to as \textit{entities} and facilitate the extraction of features in the dialog. In our chatbot, the \textit{entities} include middleboxes, SLA requirements, temporal restrictions, and endpoints targeted by the user's intent. One key advantage of using DialogFlow is the ability to deploy our chatbot across multiple platforms, including Google Assistant (present in numerous Google devices), Amazon's Alexa, or messaging apps, such as Slack and Facebook's Messenger. This feature can be helpful for a home-network user to configure her network using voice-activated assistants like Amazon's Alexa---for example, she could request parental control for her kids' devices.

Despite being extremely useful for user interactions, simply using a chatbot does not fulfill all the requirements for intent-based network planning. The \textit{entities} extracted from natural languages result in key-value pairs representing the user utterances. However, these pairs do not reflect the network configuration commands. For instance, if a network operator asks a chatbot \textit{``Please add a firewall for the backend.''}, a possible extraction result, depending on how the chatbot is built and trained, would be the following \textit{entities}: \textit{\{middleboxes: `firewall'\}}, \textit{\{target: `backend'\}}. Hence, after the chatbot interaction, we still need to translate the \textit{entities} into a structured intent that can be implemented in a destination network.

\subsection{Intent Translation}
\label{sec:translation}

In DialogFlow, after the chatbot interface extracts all the required \textit{entities} from the user utterances,  the framework calls a Rest API in a backend service designated by a WebHook, which allows us to perform the heavy processing for translations.  We configured a WebHook from our chatbot to our Intent Translator to receive all the extracted \textit{entities}. These \textit{entities} are fed to a previously trained sequence-to-sequence learning model \cite{Sutskever2014},  which translates \textit{entities} to structured intents written in our \textit{Nile} language (detailed in \textsection\ref{sec:language}). 

\begin{figure}[!ht]
	\centering
	\includegraphics[width=\linewidth]{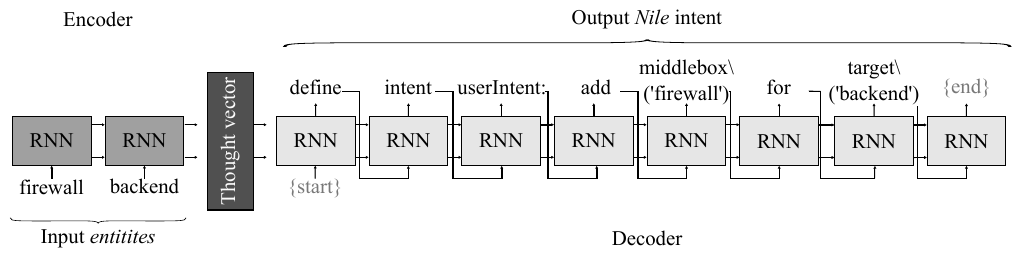}
    \caption{Sequence-to-sequence learning model.}
    \label{fig:seq2seq}
\end{figure}

A neural sequence-to-sequence learning model consists of two Recurrent Neural Network (RNN) with Long Short-Term Memory (LSTM) hidden units: an \textit{encoder} and a \textit{decoder}. In this model, the RNN \textit{encoder} processes the sequence of words (in our case the extracted \textit{entities}) and generates a thought vector, which is a numerical representation of the input sequence. The RNN \textit{decoder} receives the thought vector as input and generates a sequence of words in the destination language (in our case \textit{Nile}). Figure~\ref{fig:seq2seq} shows an example of the encoding-decoding process. Note that the RNNs allow input and output sequences of different lengths.

One of the shortcomings of using neural networks for text-to-text translations is the enormous vocabulary that each language has, which requires large datasets and substantial time to train the models. However, as we are using previously extracted \textit{entities} as input and a limited and well-defined language as output, we can overcome the shortcoming above by performing \textit{entities' anonymization}~\cite{Iyer2017}. This pre-processing consists of replacing each extracted \textit{entity} with a token representing it and using the token representation as input for the RNN \textit{encoder}.  For example, if the Entities Extractor outputs \textit{``firewall''} and \textit{``backend''}, we would use \textit{anonymization} to convert them to the tokens \textit{`@middlebox'} and \textit{`@target'} before starting the Intent Translation stage. After the translation, we simply run a \textit{deanonymization} on the resulting intent program to replace the tokens with the originally extracted entities. By using anonymization, we can reduce the number of training cases needed for the model considerably, since we do not have to consider every possible \textit{entity} value for network intents. Our preliminary tests showed a size reduction of the training dataset from 1.000.000 to 5.000 with, surprisingly, improved accuracy.

As we cannot use words directly as input for the sequence-to-sequence model, we convert each input word of the model to a unique numerical representation.  The numerical representation of the anonymized \textit{entities} are the numeric indices in a pre-built dictionary that contains all words in the model. Equation \ref{eq:index} presents an example of a conversion using a vocabulary with just four words that include the words \textit{\@middlebox} (index 2) and \textit{\@target} (index 3).

\begin{equation}
    \small
  	\left[ 'firewall', 'backend' \right] \Rightarrow \left[ '@middlebox', '@target' \right] \Rightarrow \left[ 2, 3 \right] 
	\label{eq:index}
\end{equation}

In addition to indexing the words of the input sequences, we perform Word Embedding vectorization in the first layer of the RNN \textit{encoder} to concisely represent the indexed words as arrays of real values. This word vectorization is known to improve the learning rates and prediction accuracy of linguistic models, as it can capture and represent the meaning of each word~\cite{mikolov2013}. The array of real values, which represents the sequence of anonymized \textit{entities} given as input to the sequence-to-sequence model, is then processed one by one by the RNN \textit{encoder} to generate the thought vector. The RNN \textit{decoder} then uses the encoded thought vector to predict a sequence of statements in the output language \textit{Nile}. The structured intent definition generated by the \textit{decoder} is then presented to the network operator for confirmation on the extracted desired behavior through the chatbot interface. 

The operator may either confirm the correctness of the intent program or make adjustments if necessary. After the operator's response, the intent program and the input \textit{entities} are included in the training database of the sequence-to-sequence model, and a new training round is initiated. In this interaction, we explicitly consider the operator's feedback during the translation, ensuring that the results improve every time the operator requests an action.

\subsection{Intent Deployment}
\label{sec:deployment}

Finally, having a structured intent program verified by the operator, the Intent Deployer can compile and deploy it into a destination network, as shown in Figure~\ref{fig:architecture}.  In this stage, we make assertions to verify any conflicts between the extracted intent and the network configuration and warn the operator through the chatbot interface.  We then translate \textit{Nile} programs into configuration commands using SONATA-NFV~\cite{Peuster2016}. We currently do not deal with non-SDN networks, but we intend to develop an AI-based module that can handle different networks in future work. For example, a neural network could infer the best routes to comply with SLA requirements of the intent without the need of a pre-populated database. However, the decoupling provided by the intent definition language allows compilations to other existing network configurations, including other policy languages, such as Janus~\cite{Abhashkumar2017}, PGA~\cite{Prakash2015}, and Kinectic~\cite{Kim2015}.

Ideally, the process described in this section and presented in Figure~\ref{fig:architecture} would also include an Intent Behavior Monitor module.  This module would ensure that the deployed policies respect the intents extracted by the refinement process. To achieve this goal, the module could leverage a neural network to predict which parameters should be monitored. The module could then monitor the parameters and notify the operator in case of disparities between the behavior and the intent. We leave the design and implementation of this module for future work.

%% file: src/language.tex
\section{\textit{Nile}: Intent Definition Language}
\label{sec:language}

The previous section presented a lengthy process to transform natural language into device configurations.  A key insight we uncovered from this translation process is the clear need for a simple, yet comprehensive, abstraction layer between lower-level policies and the natural language used by operators and home users. While low-level policy enforcers, such as SDN rules, require operators with extensive expertise and management experience to program the intended behavior of a network, natural language is hard to parse and interpret correctly and often inaccurate, creating a huge gap between the intended behavior and the network configurations. Also, translating natural language intent directly to network rules decreases portability and reusability, since each possible destination network has specific features and configuration requirements. To bridge this gap, we propose the \textit{Nile}\footnote{Nile comes from Network Intent LanguagE} language as an intermediate intent representation that is close to natural language. However, \textit{Nile} exhibits enough structure that works well as the target for the learning algorithm and allows translation to different target networks.

By introducing an intent definition language as an intermediate representation in the refinement process, we decouple the policy extraction from the policy deployment and enforcement. This decoupling, with an intermediate representation that resembles natural language and is easy to understand, allows us to use the feedback from the operator before deploying the extracted behavior. Moreover, the intent definition language acts as an abstraction layer for other policy mechanisms, reducing the need for operators to learn multiple policy languages for each different type of network. Hence, the design requirements for the intent language grammar are:  \textit{(i)} high legibility, as operators unfamiliar to the language must be able to understand and assert the correctness of the intent;  \textit{(ii)} high expressiveness, to faithfully represent the operator's intention; and \textit{(iii)} high writability, to allow operators to make adjustments to the generated intents quickly and easily. The grammar of \textit{Nile}, in EBNF notation~\cite{EBNF-ISO}, is in Grammar \ref{gra:nile}.

\begin{grammarbox}
   \small
  \begin{grammar}
    <intent> ::= `define intent' \textcolor{red}{intent_name} `:' <commands> 

\vspace{-0.5mm}    
    <commands> ::= <command> \{ '\verb!\n!' <command> \}
    
\vspace{-0.5mm}    
    <command> ::= (<middleboxes> | <qos> | <rules>)+ [ <optional> ]
    
\vspace{-0.5mm}    
    <middleboxes> ::= `add' <middlebox> \{ (`,' | `, \verb!\n!') <middlebox> \} 
        
\vspace{-0.5mm}    
    <middlebox> ::= `middlebox(' \textcolor{red}{middlebox_id} ')'
    
\vspace{-0.5mm}    
    <qos> ::= `with' <metrics>
        
\vspace{-0.5mm}    
    <metrics> ::= <metric> \{ (`,' | `, \verb!\n!') <metric>\}
    
\vspace{-0.5mm}    
    <metric> ::= <metric_id>`(' <constraint> `,' \textcolor{red}{value} `)' 
    		\alt <metric_id>`(none)'
    
\vspace{-0.5mm}    
    <metric_id> ::= latency | jitter | loss | throughput
    
\vspace{-0.5mm}    
    <constraint> ::= `less [or equal]' | `more [or equal]' | `equal' | `different'
    
\vspace{-0.5mm}    
    <rules> ::= <rule> \{ `\verb!\n!' <rule> \}
    
\vspace{-0.5mm}    
    <rule> ::= (allow | block) <traffic>
    
\vspace{-0.5mm}    
    <optional> ::= <targets> | <locations> | <interval>
    
\vspace{-0.5mm}    
    <targets> ::= `for' <target> \{ (`,' | `, \verb!\n!') <target> \}
    
\vspace{-0.5mm}    
    <target> ::= `client(' \textcolor{red}{client_id} `)' | <traffic>
    
\vspace{-0.5mm}    
    <locations> ::= `from' <endpoint> `to' <endpoint>
    
\vspace{-0.5mm}    
    <endpoint> ::= `endpoint(' \textcolor{red}{endpoint_id} `)'
    
\vspace{-0.5mm}    
    <interval> ::= `start' <date_time> `\verb!\n!' `end' <date_time>
    
\vspace{-0.5mm}    
   	<traffic> ::= `traffic(' \textcolor{red}{traffic_id} `)' | `flow(' [<five_tuple>]+ `)'
    
\vspace{-0.5mm}    
    <five_tuple> ::= `protocol:' \textcolor{red}{v} 
   			| `src_port:' \textcolor{red}{v} 
            | `src_ip:' \textcolor{red}{v} 
            | `dest_port:' \textcolor{red}{v} 
            | `dest_ip:' \textcolor{red}{v}             
        
\vspace{-0.5mm}    
    <date_time> ::= `datetime('\textcolor{red}{datetime}`)' | `date('\textcolor{red}{date}`)' | `hour('\textcolor{red}{hour}`)'
  \end{grammar}

\caption{\textit{Nile}.}
  \label{gra:nile}
\end{grammarbox}

With the Nile language, we can build powerful yet simple intents. For example, an input \textit{"Add firewall and intrusion detection from gateway to backend for client B, with latency less than 10ms and 100mbps of bandwidth, and allow HTTPS only, everyday from 09:00 to 18:00 "} can be represented as in Listings \ref{ls:example2}. Note that the Nile program includes only the specific hours defined in the intent, which means that the behavior must be repeated every day. This example illustrates how \textit{Nile} provides a high-level abstraction for structured intents. We believe this initial grammar for \textit{Nile} is expressive enough to represent most network intents, but we do plan to expand it to incorporate new features. Note that the ids provided by the operator (\textit{i.e.}, tokens in red in Grammar \ref{gra:nile}) must be resolved during the compilation process, as they represent information specific to each network. This feature of the language enhances its flexibility for defining intents and serving as an abstraction layer.

\begin{lstlisting}[caption={\textit{Nile} intent example.}, captionpos=b, label=ls:example2]
define intent qosIntent:
  from  endpoint('gateway') 
  to    endpoint('database')
  for   client('B')
  add   middlebox('firewall'), middlebox('ids')
  with  latency('less', '10s'), 
        throughput('more or equal', '100mbps')
  allow traffic('https')
  start hour('09:00')
  end   hour('18:00')
\end{lstlisting}

%% file: src/implementation.tex
\section{Implementation}
\label{sec:implementation}

We use a different Github project for each stage of the refinement process so that people can download and reuse the stages individually.  We implemented the Entities Extractor as a DialogFlow chat interface and deployed it for testing in the Google Assistant. The chat interface consists of a list of entities, which are the key features to be parsed from natural language, and language intents (not related to network intents). Language intents represent possible user interactions that the chatbot creator provides for machine learning training so that DialogFlow can generalize and learn how to extract the necessary entities from future user interactions. We exported our implementation of the chat interface from DialogFlow as JSON files and uploaded them to GitHub\footnote{Available at \url{https://github.com/asjacobs92/nia-chatbot/}}. For reproduction purposes, the files can be imported to a new DialogFlow project and retrained. 

We implemented the Intent Translator as a Python Restful API service that is called by the DialogFlow chat interface right after it extracts the entities. The service can interact with the chatbot interface to ask for additional information if necessary. Besides this interaction, the API provides a sequence-to-sequence model developed using Keras ~\cite{Chollet2015} that we used to train our model.  We trained and computed the weights of our model with an automatically generated dataset of input entries containing examples of anonymized entities and the correspondent \textit{Nile} program, which is also anonymized. We used different sizes of datasets in our evaluation, and the results are in \textsection\ref{sec:evaluation}.  After generating a \textit{Nile} intent and confirming it with the user feedback, we retrain the model by adding the intent to the training dataset.  The Intent Translator project is available at GitHub\footnote{Available at \url{https://github.com/asjacobs92/nia-webhook/}}. For testing purposes, we deployed the Intent Translator using Heroku~\cite{heroku}.

Finally, we developed the Intent Deployer as a separate Python Restful API service that is called by the Intent Translator when it finishes the translation process. As a proof-of-concept, we developed a project that implements service chaining policies using  SONATA-NFV~\cite{Peuster2016} and deploys them in an emulated network. SONATA-NFV is an emulation platform based on  Mininet~\cite{Lantz2010} that deploys network functions as Docker containers. \textit{Nile} commands for client identification, time, and traffic requirements are not implemented yet. However, we do plan to extend our implementation to include the full set of \textit{Nile} commands and to introduce machine learning to predict the best way to compile and fulfill \textit{Nile} programs in a destination network.  This project is also available at Github\footnote{Available at \url{https://github.com/asjacobs92/nia-deployer/}}.

%% file: src/evaluation.tex
\section{Evaluation}
\label{sec:evaluation}

To assess the feasibility of our intent refinement process, we evaluate two main aspects: \textit{(i)} the accuracy we can achieve with different sizes of training datasets, aiming to find the optimal ratio between dataset size (which impacts the training time significantly) and prediction accuracy; and \textit{(ii)} the impact of the operator feedback on the accuracy of predictions over time to determine if it improves accuracy. Also, we provide a test case to demonstrate the end-to-end deployment process of intents in a destination network (i.e., from natural language to network configurations). We run our experiments on a server with 8 Intel(R) Core(TM) i7-6700 CPU at 3.40~GHz, 16GB of RAM, running Deepin Linux kernel 4.14. We generated the datasets automatically with random sets of \textit{entities} and \textit{Nile} intent pairs, combining a different number of middleboxes, endpoints, traffic matching rules, time, and QoS requirements in each intent. All training iterations were done with 70 epochs, batch size of 64, and a validation split of 20\%.  We evaluate five different sizes of training datasets: 100, 500, 1000, 2000, and 5000 entries. For each size of the training dataset, we generate a separate testing dataset, containing 20\% of the number of entries from the training dataset. 

\begin{figure}[!h]
	\centering
    \subfigure[Translation accuracy.]{\label{fig:accuracy}\includegraphics[width=0.49\linewidth]{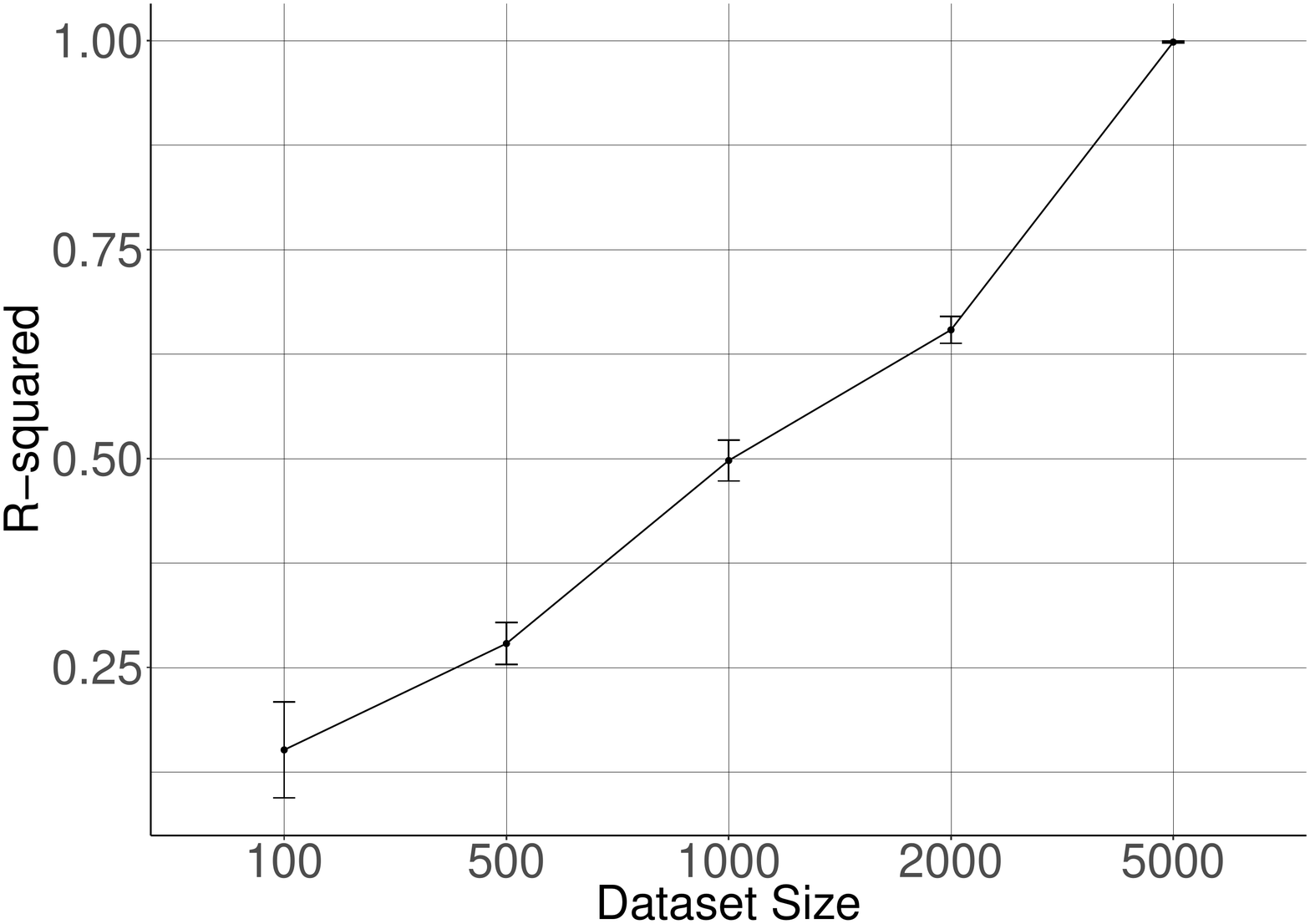}}
    \subfigure[Training time.]{\label{fig:time}\includegraphics[width=0.49\linewidth]{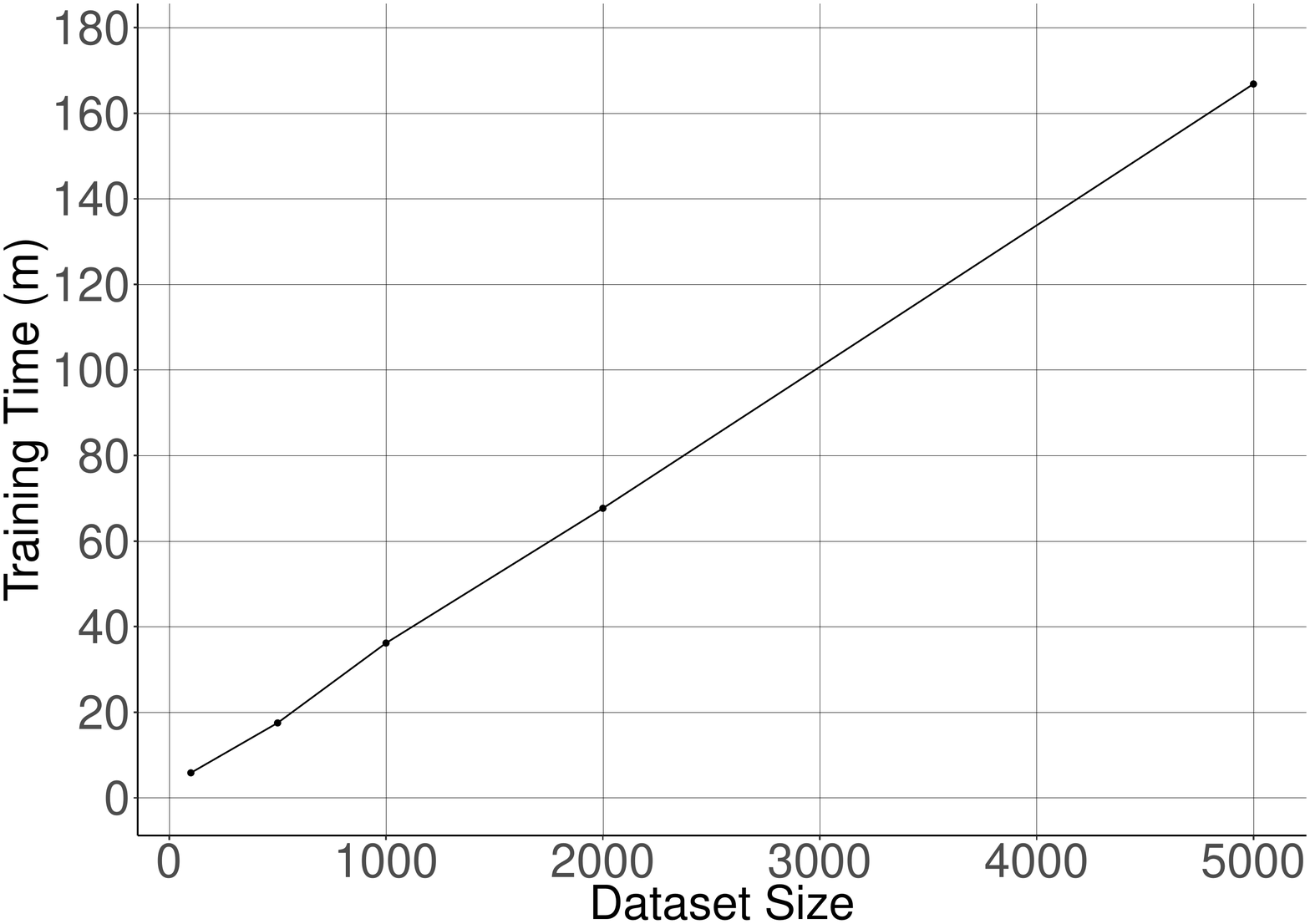}}
    \label{fig:datasets}
    \caption{Accuracy and training time by dataset sizes.}
\end{figure}

To assess the first aspect, we first train the translation model and then we measure for each prediction in the testing dataset the correlation coefficient squared (\textit{i.e.}, R-squared) between the intent predicted by the model and the expected output intent. In this case, the closer to 1 the R-squared value is, the more accurate the translation model is---i.e., less errors (\textit{e.g.}, repeated words) and wrong instructions in the resulting \textit{Nile} program. The measurements generated a list of R-squared values for each test case. Figure \ref{fig:accuracy} shows the mean and 95\% confidence interval for the measured R-squared values for each training dataset size. We also show in Figure \ref{fig:time} the training times for the same datasets. As expected, the larger the training dataset, the more accurate the results yielded by the translation model and the longer the training times. We can see from Figure \ref{fig:accuracy} that we need only 5000 entries in the training set to achieve excellent results in the refinement process. We expect better results with larger datasets. However, the training process for our largest dataset was close to three hours, and larger datasets require even longer periods for training the model. 

Next, we evaluate the impact of the operator feedback in the accuracy of the prediction with the same datasets. To simulate this process, we first load the neural network weights for the trained model. Having a trained model, we use 30 different test cases of \textit{entities} and expected \textit{Nile} intents to simulate requests from an operator.  For each case, we first use the \textit{entities} to predict an intent from the translation model, and measure the R-squared for that case in comparison to the expected \textit{Nile} intent; then, we add the expected intent into the training dataset of the model, and start a new epoch of training. 

\begin{figure}[!ht]
	\centering
	\includegraphics[width=\linewidth]{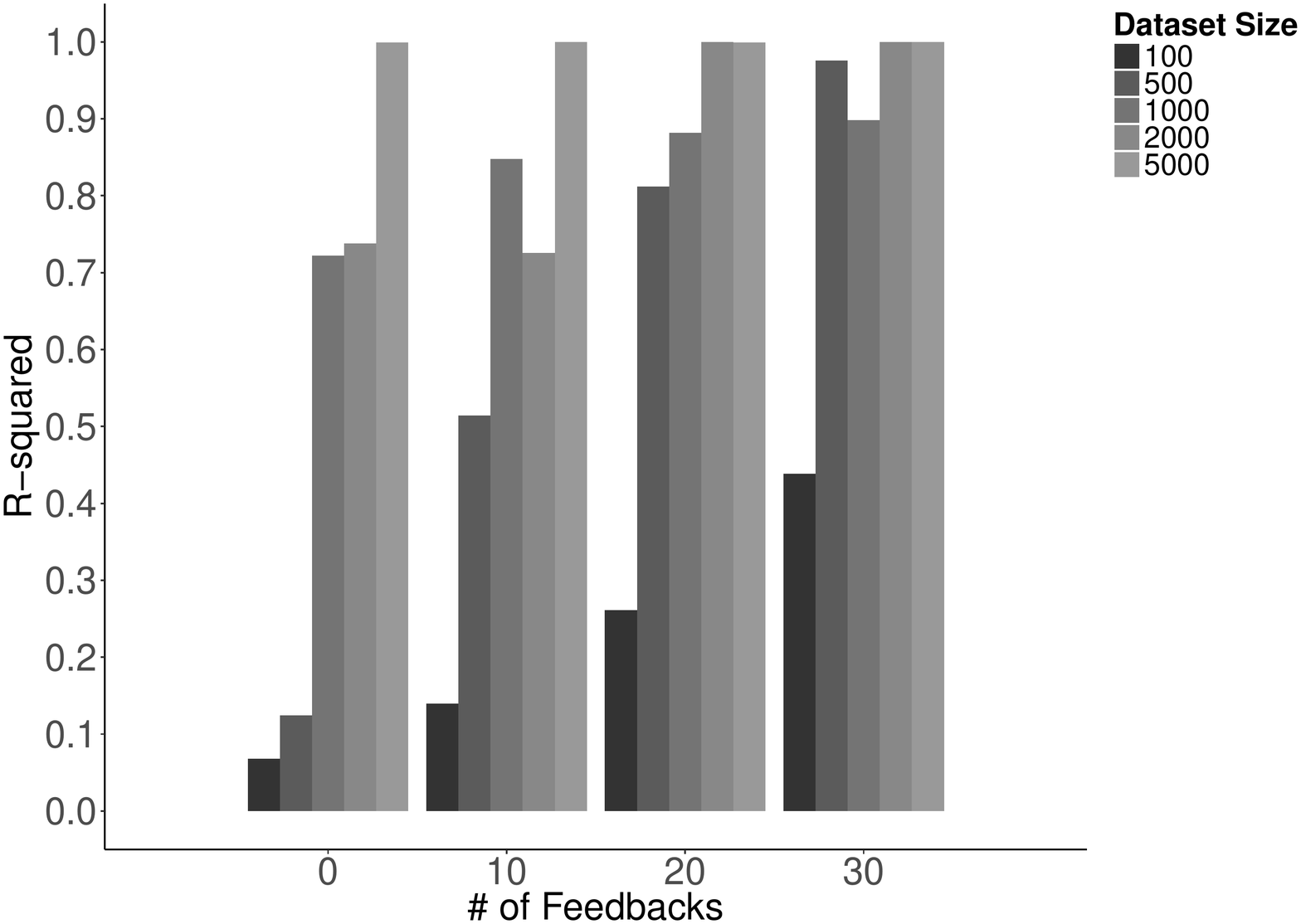}
    \caption{Accuracy improvement with feedback over time.}
    \label{fig:feedback}
\end{figure}

Figure \ref{fig:feedback} shows the R-squared values after 0, 10, 20 and 30 feedbacks were incorporated into the training dataset. It is clear from the plot that, regardless the size of the training dataset, the accuracy improves considerably with repeated training after incorporating feedback.  This behavior is particularly evident for training datasets with a smaller number of entries.  For instance, we can observe that the operator feedback can improve the accuracy of the model trained with 2000 entries up to the same level as the model trained with 5000 entries without the feedback. This result means that for a much smaller dataset, which requires much less training time, we can achieve similar results.  It is also worth mentioning that, in some cases, results obtained with smaller datasets were better than the results obtained with larger datasets, such as with dataset sizes 500 and 1000. This behavior is most likely because of feedback cases that repeated during the test since the training dataset were randomly generated. Hence, the model trained with a smaller dataset could predict with higher accuracy the cases he had already learned from the feedback.

Finally, one could argue that, since the training dataset with 5000 entries has close to perfect accuracy from the start, there is no need for incorporating feedback into the training process. We counter this argument by pointing out that, no matter how accurate the prediction model is, there will always be specific cases that are not covered by the training dataset, and the sequence-to-sequence model may produce erroneous results. Therefore, it is imperative that an operator confirms the generated intent to avoid misconfigurations, and corrects it so that the model can learn and reduce the frequency of these cases.

\begin{figure}[!h]
	\centering
    \includegraphics[width=\linewidth]{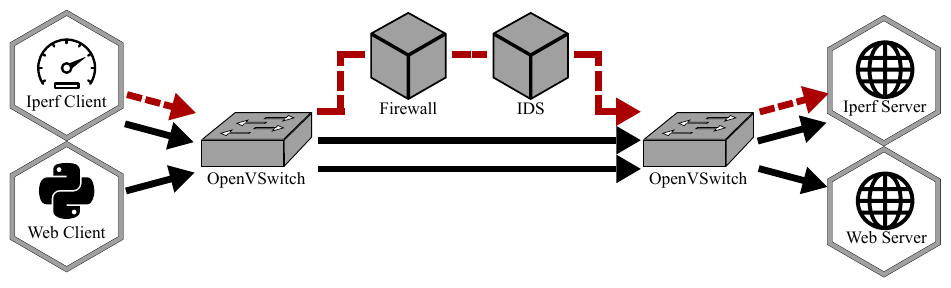}
    \caption{End-to-end test scenario.}
    \label{fig:test}
\end{figure}

To illustrate how the end-to-end deployment process of intents works,  we present a test case that concerns service chaining using SONATA-NFV (see Section 4).  All the scripts and reproduction artifacts of this and other test cases that we cannot show because of space limitations are available at Github\footnote{\url{https://github.com/asjacobs92/nia-experiment}}. The scenario consists of a network with two OpenVSwitches connecting an Iperf client that sends 100 Mbps of UDP traffic to its server and a Stratos Web client that generates HTTP requests to a Web server. After starting both clients, we tested the user intent "Please add a firewall and an IDS from Iperf client to server," aiming to block and inspect the traffic generated from the Iperf client while ignoring the traffic from the Web client.  The Entities Extractor, in DialogFlow, extracts the origin, destination and desired middleboxes of the intent, and call the Intent Translator RestAPI. The Intent Translator converts the input entities into the \textit{Nile} intent displayed in Listing \ref{ls:nile-test-case}. Subsequently, the Intent Deployer compiles the translated \textit{Nile} program in the SONATA-NFV commands shown in Listing \ref{ls:sonata-test-case}.  The middleboxes are Docker containers using pre-configured images (\textit{i.e.}, \textit{genic-vnf}) with the scripts required to run the network functions. We use iptables and Snort to implement the firewall and IDS, respectively. Figure \ref{fig:test} shows the test scenario where the red arrows represent the deployed intent.

\begin{lstlisting}[caption={Generated \textit{Nile} intent.}, captionpos=b, label=ls:nile-test-case]
define intent testIntent: 
    from endpoint('iperf client') 
    to endpoint('iperf server') 
    add middlebox('firewall'), 
        middlebox('ids')
\end{lstlisting}

\pagebreak

\begin{lstlisting}[caption={Generated SONATA-NFV commands.}, captionpos=b, label=ls:sonata-test-case]
# deploy vnfs
vim-emu compute start -d vnfs_dc -n fw \ 
  -i genic-vnf -c "./start_firewall.sh &" \
  --net"(id=in,ip=10.0.0.20/24),(id=out,ip=10.0.0.21/24)" 
vim-emu compute start -d vnfs_dc -n ids \
  -i genic-vnf -c "./start_snort.sh &" \
  --net"(id=in,ip=10.0.0.30/24),(id=out,ip=10.0.0.31/24)" 
# chain vnfs
vim-emu network add -b -src iperf-c:c-eth0 -dst fw:in
vim-emu network add -b -src fw:out -dst ids:in
vim-emu network add -b -src ids:out -dst iperf-s:s-eth0
\end{lstlisting}

%% file: src/related.tex
\section{Related Work}

Recent works on IBN feature several intent languages, frameworks, and compilers to efficiently deploy intents in network devices and middleboxes~\cite{Prakash2015, Abhashkumar2017, Sung2016, Foster2011, Anderson2014, Kim2015, Soule2014, Ryzhyk2017}.  Most notably, PGA ~\cite{Prakash2015} proposes the use of a graph abstraction to compose high-level policies and deploy them in SDN networks. PGA supports Access Control List (ACL) and service-chaining policies, leveraging a graph structure to resolve conflicts. Janus~\cite{Abhashkumar2017} extends PGA to support policies with QoS requirements, mobility, and temporal dynamics. 
More recently, Cocoon~\cite{Ryzhyk2017} introduces a framework focused on guaranteeing correctness of SDN programs that resembles our approach, but it uses first-order logic instead of machine learning to convert high-level intents into lower level configurations. Cocoon, however, does not validate its refinements with the operator or learn the operator's intent over time. Moreover, its specification language is not as user friendly as natural language.
In the industry, Robotron~\cite{Sung2016} provides a high-level intent abstraction for designing and managing the worldwide-scale network of Facebook.  While these efforts present contributions for specifying and verifying network policies, they still fail to extract intent information from pure natural language, requiring that network operators learn new and complex policy definition languages. In our work, we tackle these complexity issues by leveraging the DialogFlow chatbot interface, coupled with the neural translation process into \textit{Nile.}  By relying on \textit{Nile} solely for adjustments and confirmation, we significantly reduce the knowledge curve of our intent solution.

Other related works focus specifically on the intent and policy refinement process. For instance, INSpIRE \cite{Scheid2017} applies a refinement process to determine which middleboxes should compose a service chain to fulfill an intent. However, this refinement process focuses solely on intents related to security middleboxes, ignoring other essential complex scenarios with other intent requirements.  Machado {\em et al.}~\cite{Machado2015} and Craven {\em et al.} propose different approaches to policy refinement, leveraging Event Calculus (EC)~\cite{Machado2015} or a UML logical representation~\cite{Craven2011} as intermediate policy representations to allow the operationalization of network behavior definition.   Still, these existing research proposals for IBN fail to exploit the knowledge and feedback from the network operator. Highly complex and, sometimes, conflicting policies in network devices may cause network intents to derail from the desired behavior of the operator.  Hence, requesting feedback from the operator is crucial to avoid misconfigurations.  We tackle these shortcomings by incorporating the operator's feedback into our neural network training dataset so that we can learn from past mistakes.

%% file: src/conclusions.tex
\section{Conclusion}

In this paper, we introduced a novel intent refinement process and \textit{Nile}, a high-level intent definition language, aiming to be a step towards enabling {\em self-driving networks}. The proposed refinement process leverages a user-friendly chat interface and a sequence-to-sequence learning model that extracts from natural language a structured intent program, written in \textit{Nile}.  The extracted \textit{Nile} intent acts as an abstraction layer for lower-level configuration and policy languages, which allows us to ask for feedback from the operator before compiling the structured intent into network configurations. Our evaluation of the proposed process yielded a correlation coefficient squared (\textit{i.e.}, R-squared) of 0.99 for the intents extracted using our sequence-to-sequence model. Also, the use of feedback from the operator into the model improved the accuracy of our translation model, especially for smaller training datasets.